 \documentclass[smallabstract,smallcaptions]{dccpaper}

\usepackage{epsfig}
\usepackage{citesort}
\usepackage{amsmath}
\usepackage{amssymb}
\usepackage{color}
\usepackage{url}
\usepackage[shortcuts]{extdash}

\newlength{\figurewidth}
\newlength{\smallfigurewidth}

\begin{document}
\title
{\vspace{-10mm}
\large
\textbf{Distortion-Controlled Dithering with\\
Reduced Recompression Rate
\thanks{This material is based upon work supported by the Office of Naval Research under Contract No. N68335-21-C-0625 and the National Science Foundation Graduate Research Fellowship under Grant No. DGE-2233066.}
}
}

\author{%
Morriel Kasher$^{\ast}$, Michael Tinston$^{\dag}$, and Predrag Spasojevic$^{\ast}$\\[0.5em]
{\small\begin{minipage}{\linewidth}\begin{center}
\begin{tabular}{ccc}
$^{\ast}$Rutgers University & \hspace*{0.5in} & $^{\dag}$Expedition Technology, Inc. \\
New Brunswick, NJ, USA && Herndon, VA, USA\\
\url{morriel.kasher@rutgers.edu},&& \url{mike.tinston@exptechinc.com}\\
\url{spasojev@winlab.rutgers.edu}
\end{tabular}
\end{center}\end{minipage}}
}

\maketitle
\thispagestyle{empty}

\begin{abstract}
Dithering is a technique that can improve human perception of low-resolution data by reducing quantization artifacts.
In this work we formalize and analytically justify two metrics for quantization artifact prominence, using them to design a novel dithering method for distortion-controlled data compression. We present theoretical entropy calculations for this dither and experimentally validate its performance on a low-rate image compression task. The result is a drastic improvement in the perceptual quality of quantized images with a lower recompression entropy than any state-of-the-art dither technique, achieving 45 points lower PIQUE at the same rate or 40\% lower rate at the same PIQUE. The proposed dither is an adaptable tool applicable for use in any lossy compression system, permitting precise control of rate-distortion characteristics for both compression and recompression.
\end{abstract}

\vspace{-2mm}
\Section{Introduction}
\vspace{-2mm}
Quantization is a ubiquitous process in lossy data compression \cite{GrayQuant1998}, underlying several fundamental compression techniques \cite{BergerLSC1998}. Many theoretical results for lossy compression performance rely on an assumption of uncorrelated quantization error \cite{DerpichUncorr2008}, which is approximately true for high-rate compression as per the popular pseudo-quantization noise model \cite{WidrowQuant1996}. For low rates, quantization error can instead be made uncorrelated by the use of appropriately-implemented dithering \cite{GrayDither1993}.

The discovery of this and other useful properties of dithering has led to its successful integration in many lossy compression schemes including transform coding \cite{AkyolTransform2009}, subband coding \cite{AshourianImagesubband2002} \cite{BudsabathonAudiosubband2006}, and entropy coding \cite{ZamirECDQ1995}. Today dithered quantization is prevalent in a variety of applications including image processing (halftoning) \cite{LauHalftoning2018}, digital watermarking \cite{EggersWatermarking2000}, and federated learning \cite{GandikotaGradient2022}.

Despite this, relatively little attention has been devoted to advancing the theoretical basis of dithering for compression beyond some initial entropy considerations \cite{SchobbenDitheredcompression1997}. Among the most promising prospects is the application of dither to near-lossless compression \cite{KlimeshNLC2000} and the closely-related idea of distortion-controlled compression \cite{KlimeshDistortion1999}. While these prior works have shown the potential of dithering, they lack a mathematically-rigorous justification for its usage, do not provide practical tools for dither design, do not compare performance of different dither distributions, and ultimately fail to clarify when dithering should be used at all.

Our work addresses these shortcomings in several ways. First, we hypothesize a relationship between the perceptual prominence of quantization artifacts and the autocorrelation properties of the quantization error. From this we analytically derive two new formulations for the distortion-control problem using novel error metrics. Next, we propose a set of parametric dither distributions adaptable for a desired distortion criterion. Finally, these distributions are compared theoretically and experimentally for both rate and distortion, in the process validating our previous hypothesis.


\Section{Background}
\SubSection{Quantization and Dithering Theory}
Consider a quantization operation $Q(.)$ defined on analog inputs $w \in \mathbb{R}$. For such a finite-level $b$-bit quantizer, let there be $2^{b}$ unique monotonically-increasing digital codebook values given by the vector $\bf{C}$ and $2^{b}+1$ unique monotonically-increasing analog partition levels given by the threshold vector $\bf{T}$ such that:
\begin{equation}
    Q(w) = C_{k} \text{ iff } T_{k} < w \leq T_{k+1}, k = \{1, \hdots, 2^{b}\}
\end{equation}
A quantizer is \textit{uniform} if $C_{k+1}-C_{k} = T_{k+1}-T_{k} = \Delta$ for $k = \left\{2, \hdots, 2^{b}-1\right\}$.
An infinite ($k \in \mathbb{Z}$) uniform quantizer can either be \textit{mid-tread} ($C_{k} = k\Delta, T_{k} = k\Delta - \Delta/2$) or \textit{mid-riser} ($C_{k} = k\Delta + \Delta/2, T_{k} = k\Delta$).

A \textit{dithered} quantizer forms the quantizer input $w$ as a sum of an analog input sample $x \in X$ and an additive dither sample $v \in V$ such that $w = x+v$. 
One can either use \textit{non-subtractive dither} (NSD) with output $y = Q(x+v)$ or \textit{subtractive dither} (SD) with output $y = Q(x+v)-v$. 
Quantization error 
$\varepsilon \triangleq y - x$ can be written as $\varepsilon_{\text{NS}} = Q(x+v)-x$ for NSD and $\varepsilon_{\text{S}} = Q(x+v)-(x+v)$ for SD.

The choice of distribution $f_{V}(v)$ used to generate the random dither variable $V$ is non-trivial. When no dithering is used ($f_{V}(v) = \delta (v)$) the quantization error is highly correlated with the input signal $\mathbb{E}[x \varepsilon_{\text{NS}}] \neq 0$. This results in undesirable artifacts such as spectral harmonic spurs, step-like boundaries within the signal, and other prominent perceptual features. However, without dithering the Mean Square Error (MSE) is minimized to $\mathbb{E}[\varepsilon_{\text{NS}}^{2}] = \Delta^{2}/12$. We define full dithering as when the dither PDF is of the form $f_{V}(v) = \Pi_{\Delta}(v)$ where:
\begin{align}
    \Pi_{a}(v) \triangleq 
    \left\{\begin{matrix}
    \frac{1}{a},& -\frac{a}{2} \leq v \leq \frac{a}{2}\\ 
    0,& \textup{otherwise}
\end{matrix}\right.
\end{align}

Full dithering increases the MSE $\mathbb{E}[\varepsilon_{\text{NS}}^{2}]$ by 3 dB to $\Delta^{2}/6$ but makes the quantization noise white ($\forall k \neq 0, \mathbb{E}[\varepsilon_{\text{NS},0}\varepsilon_{\text{NS},k}] = 0$) with a first conditional error moment $\mathbb{E}[\varepsilon_{\text{NS}} | x] = 0$ \cite{WannamakerNSD2000}. When subtractive, full dithering also has the advantage of making the error $\varepsilon_{\text{S}}$ statistically independent of the input and uniformly distributed on $[-\Delta/2, \Delta/2]$ with no increase in MSE as $\mathbb{E}[\varepsilon_{\text{S}}^{2}] = \Delta^{2}/12$ \cite{WannamakerDither1997}. While SD outperforms NSD, it requires that the output $y$ be represented in high-resolution and necessitates a coherent (synchronized) digital estimate of the analog dither signal.
NSD avoids these issues as an exclusively source-coding method, but increases MSE and can only ensure independence of specific error moments from the input.

\Section{Analysis}

\SubSection{Motivation for Dither Optimization}
Quantizer distortion 
is classically expressed using MSE
which fails to capture the perceptual impact of 
error correlation. This motivates an alternative distortion metric encompassing both MSE (for non-perceptual applications like data communication) and some measure of error correlation (for perceptual applications like image and audio processing). Such a metric 
can then be used as a tool to design dither distributions that optimize between these two evaluation criteria. 

Distortion alone is insufficient to characterize the effect of quantization. The compression rate should also be considered, and is bounded by the information entropy of the quantized data. This motivates a dithering method which can not only optimize for the unique distortion metric outlined above, but also maintain a low  entropy value when used as a source-coding method (NSD). In many applications, data is decompressed (undergoing SD) but later must be recompressed for transmission to another user or storage elsewhere. Therefore both NSD entropy (for compression) and SD entropy (for recompression) should be reduced.

\SubSection{Distortion-Control Optimization Formulation}
Each choice of dither distribution $f_{V}$ incurs some cost (denoted $C(f_{V}$)) while reducing some measure of highly-correlated quantization distortion or artifact prominence (denoted $A(f_{V})$). The goal of distortion-controlled quantization (as defined by \cite{KlimeshDistortion1999} in an alternative form) is to choose a dither distribution that optimizes the trade-off:
\begin{equation}\label{eq:lambdaoptimizationproblem}
    \min_{f_{V}} (1-\lambda) C(f_{V}) + \lambda A(f_{V})
\end{equation}
Where $\lambda$ is a relative weight for the convex combination of the two competing objectives. 
One appropriate choice for $C(f_{V})$ is the mean square quantization error:
\begin{equation}\label{eq:costdefinition}
    C(f_{V}) = \mathbb{E}[\varepsilon_{\text{NS}}^{2}] = \frac{1}{\Delta} \int_{-\Delta/2}^{\Delta/2} \sum_{k=1}^{2^{b}} (C_{k}-x)^{2} \cdot \left[ \int_{T_{k}-x}^{T_{k+1}-x} f_{V}(v)dv \right] dx
    \triangleq \text{MSE}
\end{equation}
The choice of $A(f_{V})$ is more difficult since this must quantify the perceptual concept of quantization artifact prominence. Psychoacoustic studies have indicated that the use of dithering improves perception of quantized signals by whitening the quantization noise \cite{WannamakerPsychoacoustics1992}. Since white noise is defined as being temporally uncorrelated, we hypothesize that the perceptual prominence of quantization artifacts is proportional to the magnitude of the quantization error autocorrelation vector. 
For error values at a specific sample index $i$ given by $\varepsilon_{\text{NS},i}$ we define this
vector for $N$ lag values $\boldsymbol{r_{\varepsilon \varepsilon}} \in \mathbb{R}^{N}$ 
as $\boldsymbol{r_{\varepsilon \varepsilon}}(n, i) = \mathbb{E}[\varepsilon_{\text{NS},i}\varepsilon_{\text{NS},i+n}], n \in \left\{1, \hdots, N\right\}$. Assuming a wide-sense stationary error process ($\forall i, \boldsymbol{r_{\varepsilon \varepsilon}}(n, i) = \boldsymbol{r_{\varepsilon \varepsilon}}(n)$) allows two candidate formulations:
\begin{align}\label{eq:artifelldefn}
    A_{R_{1}}(f_{V}) = \frac{1}{N} \| \boldsymbol{r_{\varepsilon \varepsilon}}(n) \|_{1} &= \frac{1}{N}\sum_{n=1}^{N} \bigl|\mathbb{E}[\varepsilon_{\text{NS},i}\varepsilon_{\text{NS},i+n}]\bigr|\nonumber\\
    A_{R_{2}}(f_{V}) = \frac{1}{\sqrt{N}} \| \boldsymbol{r_{\varepsilon \varepsilon}}(n) \|_{2} &= \sqrt{ \frac{1}{N} \sum_{n=1}^{N} (\mathbb{E}[\varepsilon_{\text{NS},i}\varepsilon_{\text{NS},i+n}])^{2}}
\end{align}
Intuitively, $A_{R_{1}}$ represents the mean absolute autocorrelation of the quantization error, while $A_{R_{2}}$ represents the root mean square autocorrelation of the same.

\SubSection{Distortion Bounds}
With no prior information about the input signal $x$ we can bound the expressions in (\ref{eq:artifelldefn}) by assuming uniform quantization of a uniform input distribution $f_{X} = \Pi_{2^{b}\Delta}(x)$ where each dither sample $v \sim f_{V}$ is independent and identically distributed (iid). Then by applying Jensen's Inequality 
we have:
\begin{align}\label{eq:artifactdefinition}
    A_{R_{1}}(f_{V})
    \leq (\mathbb{E}_{X}[|\mathbb{E}_{V}[\varepsilon_{\text{NS}}]|])^{2} = \left(\frac{1}{\Delta}\int_{-\Delta/2}^{\Delta/2} \biggl| \mathbb{E}[\varepsilon_{\text{NS}} | x ] \biggr| dx\right)^{2} &= A_{1}(f_{V}) \triangleq \text{MACE}^{2}\nonumber\\
    A_{R_{2}}(f_{V})
    \leq \mathbb{E}_{X}[\mathbb{E}_{V}^{2}[\varepsilon_{\text{NS}}]] = \frac{1}{\Delta}\int_{-\Delta/2}^{\Delta/2} (\mathbb{E}[\varepsilon_{\text{NS}} | x ] )^{2} dx &= A_{2}(f_{V}) \triangleq \text{MSCE}
\end{align}
where we know:
\begin{equation}
    \mathbb{E}[\varepsilon_{\text{NS}} | x] = \mathbb{E}_{V}[\varepsilon_{\text{NS}}] = \sum_{k=1}^{2^{b}} (C_{k}-x) \cdot \int_{T_{k}-x}^{T_{k+1}-x} f_{V}(v) dv
\end{equation}
Terms $A_{1}$ and $A_{2}$ are key figures of merit in evaluating the performance of dithered quantizers, and we assign them the names Mean Absolute Conditional Error Squared (MACE$^{2}$) and Mean Square Conditional Error (MSCE) respectively. Likewise the cost function $C$ is the MSE. By minimizing $A_{1}$ or $A_{2}$ we can minimize an upper bound on the norm of the error autocorrelation vector in the $\ell_{1}$ or $\ell_{2}$ sense, respectively.

We propose two parametric dither distributions
(found through computational analysis of numerical simplex optimization results) for solving (\ref{eq:lambdaoptimizationproblem}) when $C(f_{V})$ is given by (\ref{eq:costdefinition}) and $A(f_{V})$ is either $A_{1}$ or $A_{2}$ in (\ref{eq:artifactdefinition}):
\begin{align}\label{eq:fvdefns}
    f_{V_{1,\alpha}}(v) &= \alpha \Pi_{\alpha \Delta}(v) + (1-\alpha)\frac{1}{2}\left[\delta \left(v - \frac{\alpha\Delta}{2}\right) + \delta \left(v + \frac{\alpha\Delta}{2}\right)\right]\nonumber\\
    f_{V_{2,\alpha}}(v) &= \Pi_{\alpha \Delta}(v)
\end{align}
Note that $f_{V_{2,\alpha}}$ was originally proposed by Klimesh \cite{KlimeshDistortion1999} whose problem formulation utilized a modified form of $A_{2}$ (albeit without rigorous justification).

We can then express our optimization objectives in terms of this parametric dither distribution as: 
$C_{m}(\alpha) = C(f_{V_{m,\alpha}})$ and $A_{p,m}(\alpha) = A_{p}(f_{V_{m,\alpha}})$
where $m = \{1, 2\}$ selects a dither distribution from (\ref{eq:fvdefns}) while $p = \{1, 2\}$ selects a problem formulation from (\ref{eq:artifactdefinition}). Direct computation of these functions is given in Table~\ref{tab:alphaobjvalues}.
\begin{table}[t]
    \centering
    \caption{\label{tab:alphaobjvalues} Parametric Optimization Objective Values (Units Normalized to $\Delta^{2}$)}
    {
    \renewcommand{\baselinestretch}{1}\footnotesize
    \begin{tabular}{|c|c|c|c|}
    \cline{2-4}
    \multicolumn{1}{c|}{~}
       & $C_{m}(\alpha) \,\,\, [\text{MSE}]$ & $A_{1,m}(\alpha) \,\,\, [\text{MACE}^{2}]$ & $A_{2,m}(\alpha) \,\,\, [\text{MSCE}]$ \\
    \hline
       $m=1$ & $(1/12) (-2\alpha^{3} +3 \alpha^{2} + 1)$ & $(1/16) (1-\alpha)^{4}$ & $(1/12) (1-\alpha)^{3}$ \\
       $m=2$ & $(1/12) (\alpha^{2} + 1)$ & $(1/16) (1-\alpha)^{2}$ & $(1/12) (1-\alpha)^{2}$ \\
    \hline
    \end{tabular}}
\end{table}

Using these expressions we can choose the optimal value of $\alpha$ that solves (\ref{eq:lambdaoptimizationproblem}) for a given $\lambda$ when using the dither distributions defined by (\ref{eq:fvdefns}). Specifically:
\begin{equation}\label{eq:alphastaropt}
    \alpha_{p,m}^{*} (\lambda) = \arg \min_{\alpha} (1-\lambda) C_{m}(\alpha) + \lambda A_{p,m}(\alpha)
\end{equation}
The solutions to (\ref{eq:alphastaropt}) are given in Table~\ref{tab:alphastarvals}.
\begin{table}[t]
    \centering
    \caption{\label{tab:alphastarvals} Optimal $\alpha$ Values for Distortion-Controlled Dithering}
    {
    \renewcommand{\baselinestretch}{1}\footnotesize
    \begin{tabular}{|c|c|c|}
    \cline{2-3}
    \multicolumn{1}{c|}{~} &
    \multicolumn{2}{c|}{$\alpha_{p,m}^{*}$}\\
    \cline{2-3}
       \multicolumn{1}{c|}{~} & $p = 1$ & $p = 2$ \\
    \hline
       $m = 1$ & $(1-\sqrt{1-\lambda^{2}})/\lambda$ & $\lambda/(2-\lambda)$ \\
       $m = 2$ & $3\lambda/(4-\lambda)$ & $\lambda$ \\
    \hline
    \end{tabular}}
\end{table}

\begin{figure}[t]
    \centering
    \includegraphics[width=\textwidth]{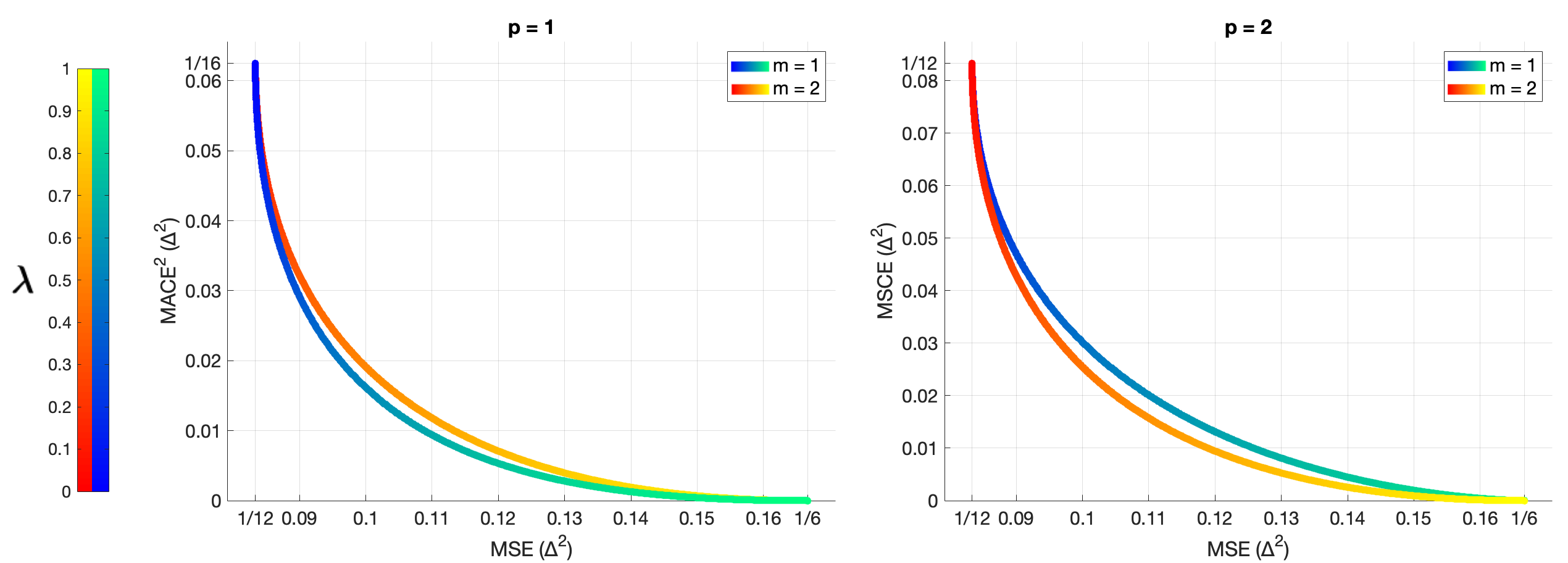}
    \caption{Analytical Multi-Objective Performance Comparison of $f_{V_{m, \alpha_{p,m}^{*}}}$ Over $\lambda$}
    \label{fig:msemscemace2curves}
\end{figure}

Fig.~\ref{fig:msemscemace2curves} plots the achieved Pareto front of the multi-objective optimization function in (\ref{eq:lambdaoptimizationproblem}) directly. From them it is clear that the best performance is achieved when $m = p$, meaning that the choice of problem formulation directly determines the best dither distribution for that problem.
Hence for convenience we use $f_{V_{m,\alpha^{*}}} = f_{V_{m,\alpha^{*}_{m,m}}}$.

\SubSection{Rate Computation}
To determine which of the two formulations in (\ref{eq:artifactdefinition}) is more relevant for compression, we analyze the output entropy after dithered quantization using each of the two proposed dither distributions in (\ref{eq:fvdefns}), as the entropy represents a lower bound on the achievable lossless encoding rate. The entropy $H$ of a random variable $Z$ after quantization is defined in bits per sample (bps) as:
\begin{equation}\label{eq:entropydefn}
    H(Q(Z)) \triangleq -\sum_{k=-\infty}^{\infty} P(Q(Z) = C_{k}) \cdot \log_{2}(P(Q(Z) = C_{k}))
\end{equation}
where the probability that a sample $z$ of the random variable $Z$ is quantized to the codebook value $C_{k}$ is given as $P(Q(Z) = C_{k}) = P(T_{k} < Z < T_{k+1}) = \int_{T_{k}}^{T_{k+1}} f_{Z}(z) dz$. We will analytically evaluate this entropy metric for a uniform random variable $X \sim U(-u/2, u/2)$ dithered by an iid random variable $V \sim f_{V_{m,\alpha^{*}}}$ and quantized by a uniform, infinite, mid-tread quantizer with width $\Delta$. This is chosen to isolate the effect of quantization from the effect of saturation due to finite levels.

When using NSD, the quantizer input $W = X+V$ has distribution $p_{W}(w) = [p_{X} \ast p_{V}](w)$, where $\ast$ denotes the convolution operation. This allows us to compute entropy using (\ref{eq:entropydefn}) directly. However when using SD, the quantizer output variable $Y=Q(X+V)-V$ has distribution:
\begin{equation}
    f_{Y}(y) = \sum_{k=-\infty}^{\infty} f_{V}(C_{k}-y) \cdot \left[ \int_{T_{k} - C_{k} + y}^{T_{k+1} - C_{k} + y} f_{X}(x)dx \right]
\end{equation}
In data compression this nominally-analog output value must be represented in some digital (higher) resolution. We consider a coarse quantization to $b$ bits $Q(.)$ (where $b < 8$) followed later by a re-quantization to 8-bit resolution with 256 levels denoted $Q_{256}(.)$. 
This is implemented for our analysis by computing the entropy of the output variable $Y$ with a new ratio $u' / \Delta = 256$ but still with infinite levels (hence some entropy values $>8$ bps). We consider this the \textit{recompression entropy} of the data.

\begin{figure}[t]
    \centering
    \includegraphics[width=\textwidth]{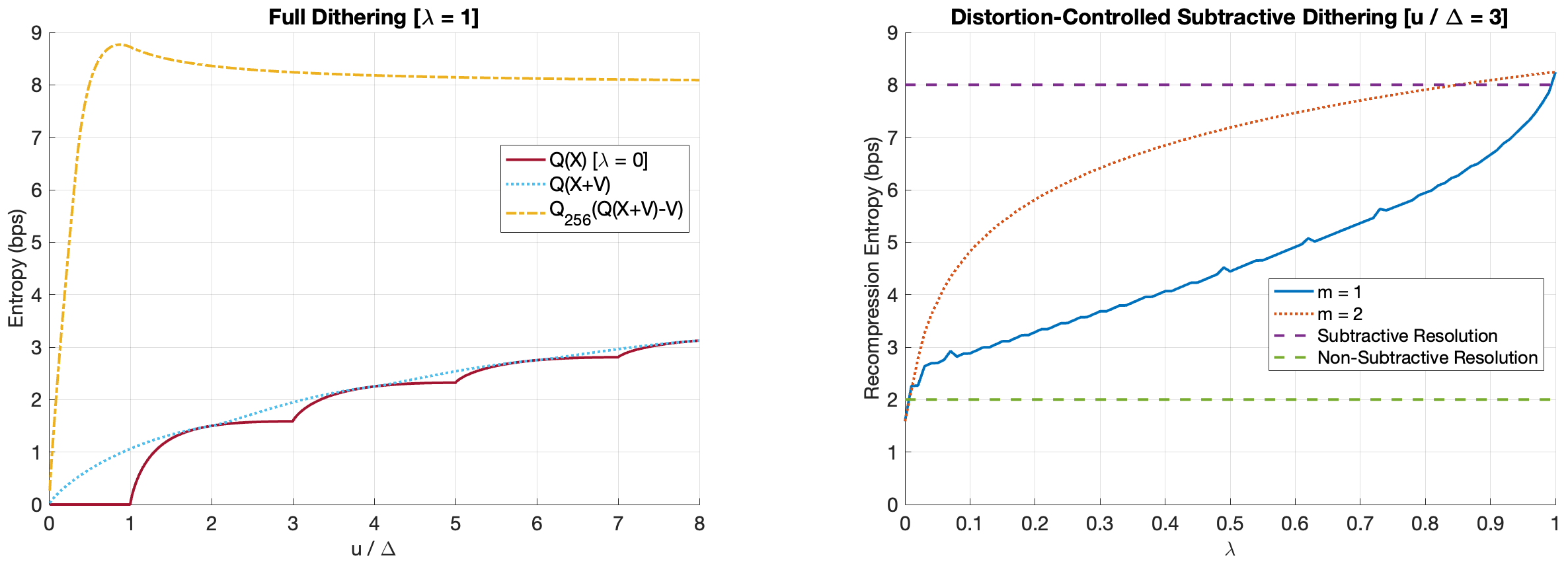}
    \caption{(a) Entropy of dithered quantized uniform distribution 
    as a function of resolution (b) Recompression entropy of subtractively-dithered 2-bit quantized uniform distribution using $f_{V_{m,\alpha^{*}}}$ as a function of $\lambda$}
    \label{fig:analyticalentropycurves}
\end{figure}

Fig.~\ref{fig:analyticalentropycurves}a plots these entropy values
as a function of a mid-tread quantization's effective resolution, computed as the ratio $u/\Delta$ of the input distribution width $u$ and the quantization level width $\Delta$ (as per \cite{SchobbenDitheredcompression1997}). While $H(Q(X+V)) \geq H(Q(X))$ at each resolution (NSD), the difference is relatively small compared to the effect of SD which almost always requires full 8-bit precision to encode $H(Q(X+V)-V)$ regardless of original quantization resolution. This establishes a motivation for partial dithering $\lambda < 1$ to mitigate this entropy requirement. 

Fig.~\ref{fig:analyticalentropycurves}b demonstrates the effect of distortion-controlled dithering using the distributions in (\ref{eq:fvdefns}) with the optimal $\alpha$ values in (\ref{tab:alphastarvals}) on the recompression entropy $H(Q_{256}(Q(X+V)-V))$ for a 2-bit quantization following by 8-bit requantization. A $u/\Delta$ value of 3 is chosen for 2-bit quantization as after dithering with up to $\Delta/2$ amplitude the overall number of levels reached by $X+V$ is 4 ($b = 2$). The results demonstrate that the dither optimized for the $\ell_{1}$ artifact function formulation ($m=1$) results in a substantially reduced recompression entropy at all $\lambda$ values compared to the $\ell_{2}$-optimized ($m=2$) dither distribution. 

\Section{Experimental Results}
Although the analytical results establish both the relevance of distortion-controlled dithering ($0 < \lambda < 1$) and the recompression entropy advantage of the proposed $\ell_{1}$-optimized dither distribution, it still remains to be seen how they perform on an example application such as image compression.

\SubSection{Image Evaluation Setup}
The $512 \times 512$ pixel test image Lena is used to evaluate our dithering technique. 
The image is natively stored as a \texttt{uint8} unsigned 8-bit integer, allowing us to compute its entropy directly using (\ref{eq:entropydefn}) with $2^8 = 256$ quantization levels. Distortion evaluation is more nuanced.
Although PSNR is a conventional image quality metric, it is strictly dependent on MSE which we already know increases monotonically over $\lambda$ without capturing the desired perceptual impact of quantization artifacts. While the popular SSIM 
metric is closer to a perceptual evaluation, it suffers from various idiosyncrasies \cite{nilsson2020understanding} which are particularly evident in coarsely quantized images. 
Instead, we adopt the Perception-based Image QUality Evaluator (PIQUE) \cite{venkata2015pique} as our metric since it does not rely on a training dataset, subjective human scores, or any reference image, while still measuring the intuitive concept of human perceptual image quality. 
PIQUE is bounded on $[0, 100]$ with a lower score indicating better image quality.

At a close viewing distance the individual pixels of an image can be ascertained, but at a farther distance several pixels in a region are perceptually blurred to the distant observer. To emulate the effect of viewing our images at 
a far distance we also evaluate the PIQUE for 
the image filtered with a circular-averaging low-pass (pillbox/disk) filter of radius one pixel. We denote this filtering operation $F[.]$.

\begin{figure}[ht]
    \centering
    \includegraphics[width=\textwidth]{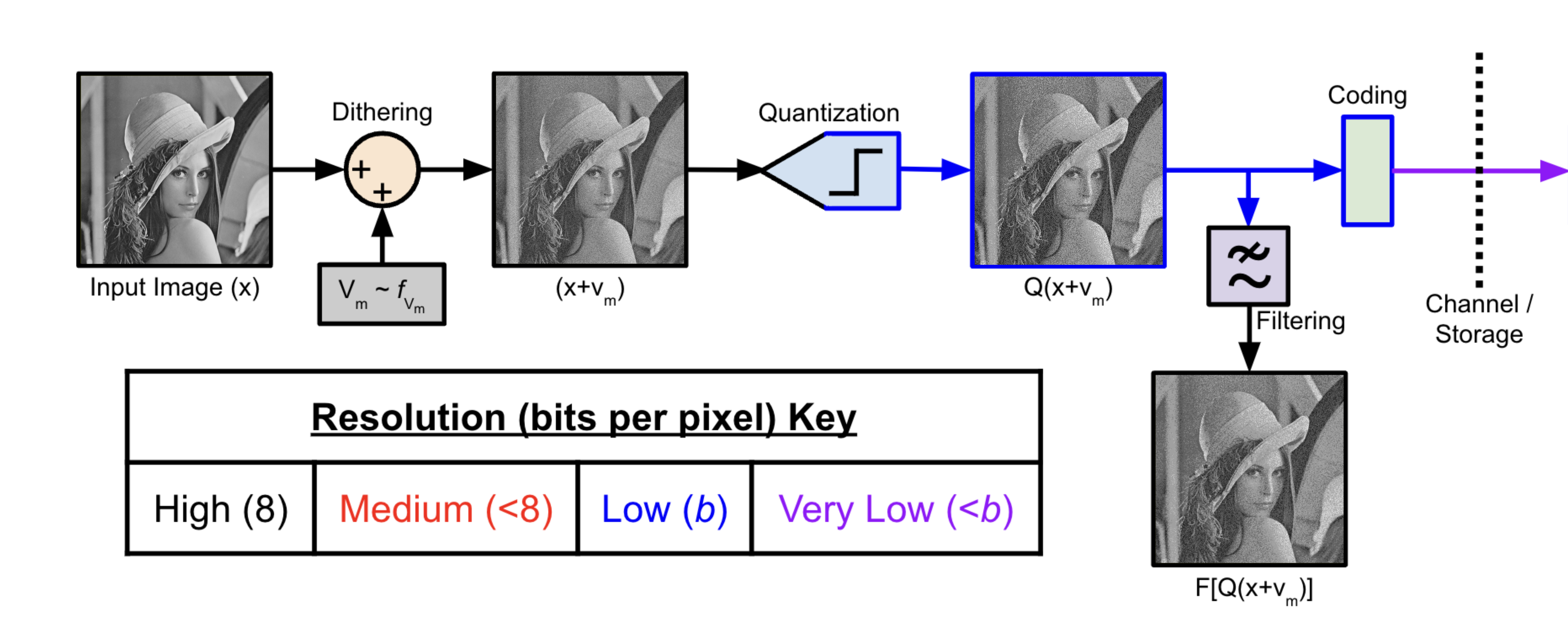}
    \caption{Image Processing Pipeline: Source Coding (Zoom for Pixel-Level Detail)}
    \label{fig:imagepipeline1}
\end{figure}

\begin{figure}[ht]
    \centering
    \includegraphics[width=0.8\textwidth]{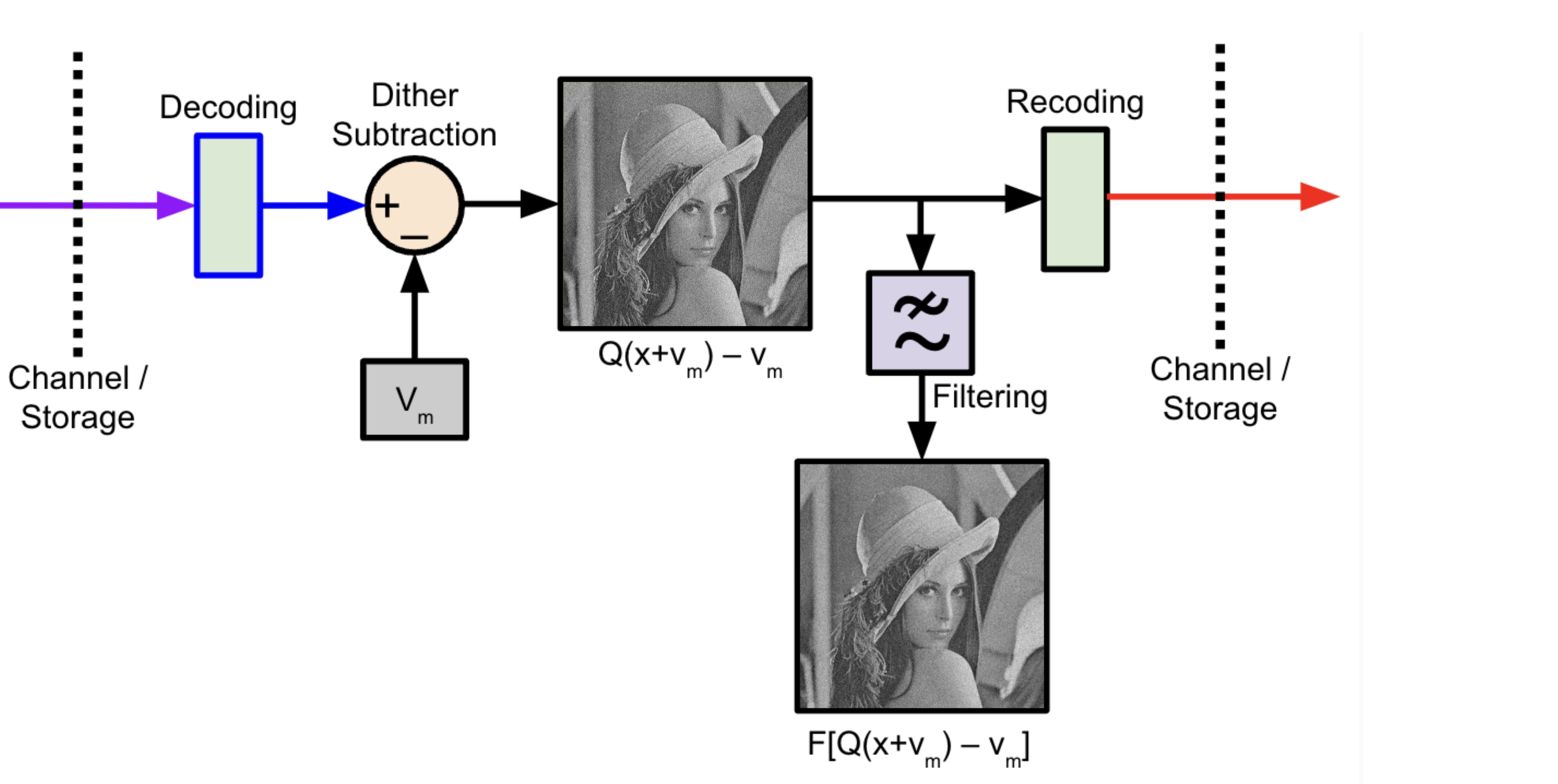}
    \caption{Image Processing Pipeline: Decoding and Recompression (Continued from Fig.~\ref{fig:imagepipeline1})}
    \label{fig:imagepipeline2}
\end{figure}

Fig.~\ref{fig:imagepipeline1} illustrates our image processing pipeline in which the input signal is rescaled by a factor of $(1 - \Delta/2)$ to avoid saturation, dithered, quantized with a uniform finite-level mid-riser quantizer, and encoded at the source. Here the NSD entropy, PIQUE (close-range), and filtered image PIQUE (far-range) are computed. Fig.~\ref{fig:imagepipeline2} depicts the decoding process including dither subtraction after which the same metrics are computed for SD and recompression.

\SubSection{Simulation Outcome}

\begin{figure}[h!]
    \centering
    \includegraphics[width=\textwidth]{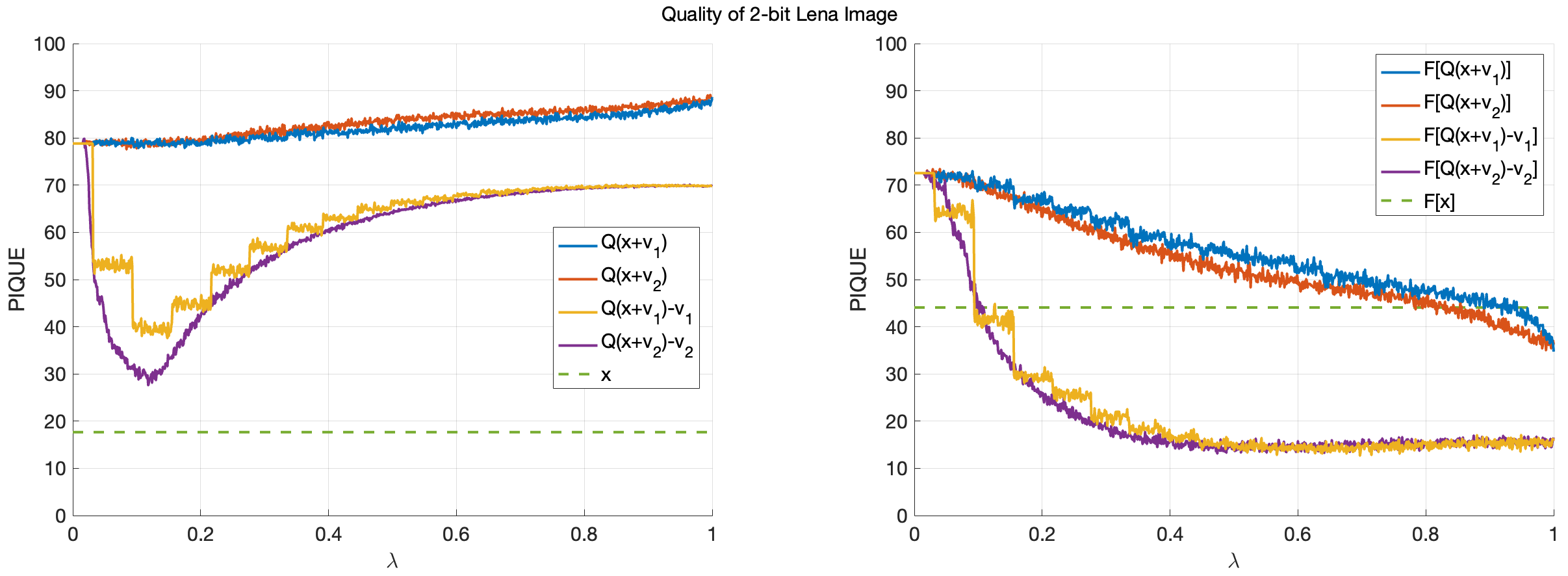}
    \caption{Experimental Image Quality over $\lambda$ Evaluated at (a) Close-Range (b) Far-Range}
    \label{fig:lenaquality}
\end{figure}
\begin{figure}[ht]
    \centering
    \includegraphics[width=\textwidth]{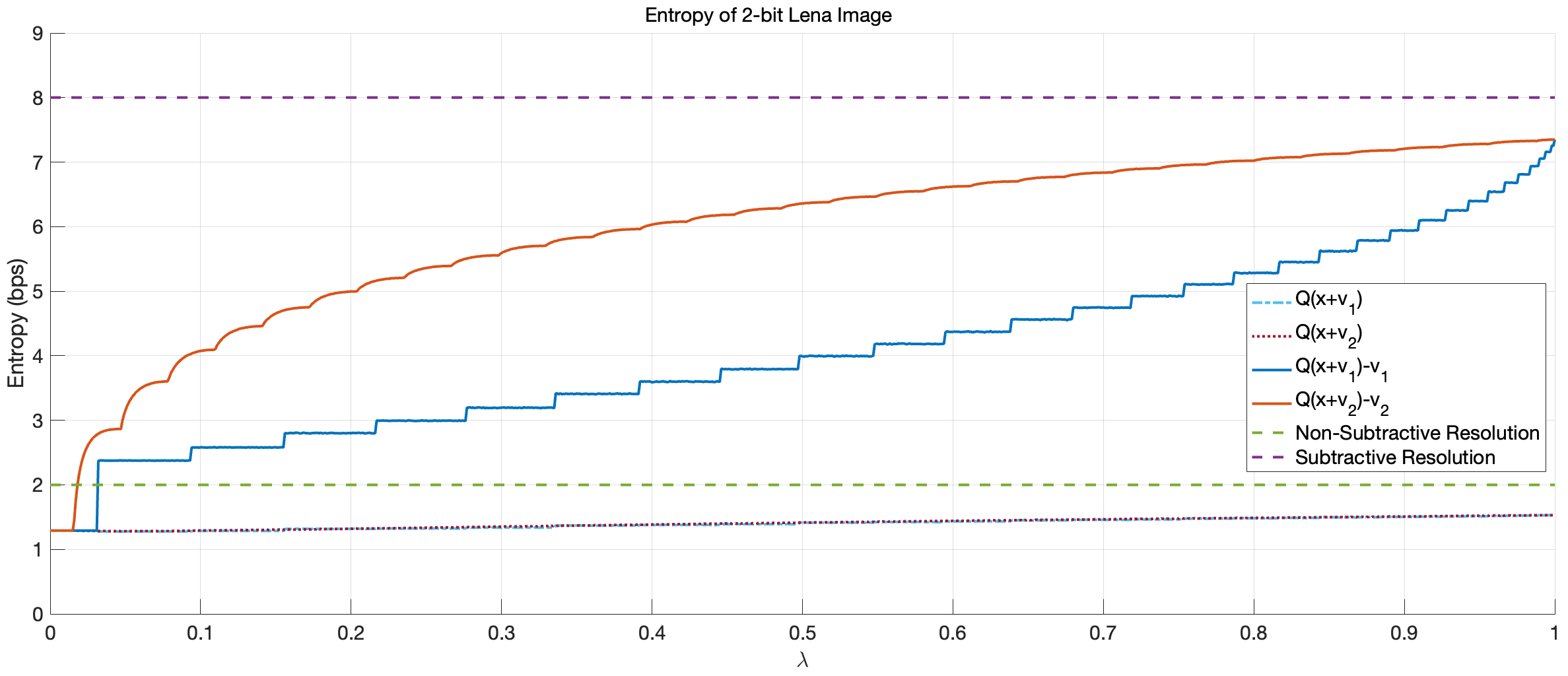}
    \caption{Experimental Image Entropy over $\lambda$}
    \label{fig:lenaentropy}
\end{figure}
\begin{figure}[h!]
    \centering
    \includegraphics[width=\textwidth]{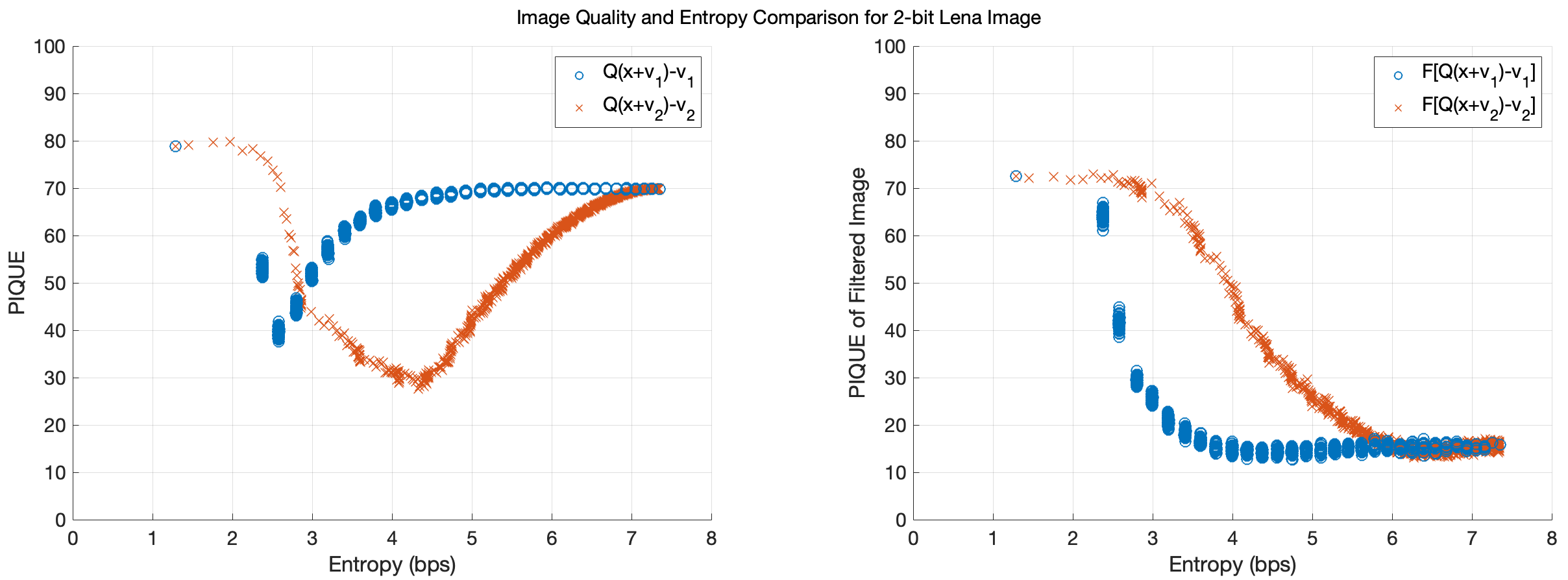}
    \caption{Experimental Image Quality as a Function of Entropy Evaluated at (a) Close\=/Range (b) Far-Range}
    \label{fig:lenaqualityentropy}
\end{figure}

Fig.~\ref{fig:lenaquality} reveals that for NSD, close-range PIQUE increases slightly over $\lambda$ while far-range PIQUE gradually decreases over $\lambda$ as the filter averages local error magnitude toward zero. With SD the close-range performance is optimized by a partial dither ($\lambda^{*} \approx 0.12$), significantly outperforming both no dithering and full dithering. At far-range the best PIQUE performance is achieved with any $\lambda \geq 0.45$, requiring much less than full dithering. Both formulations perform similarly in all cases. 

Fig.~\ref{fig:lenaentropy} shows where the two formulations diverge: the recompression entropy of the SD image. While the entropy of the NSD image is roughly constant over $\lambda$ and almost exactly equal for both formulations, the SD image requires much less entropy to re-encode when using the $\ell_{1}$-optimized dither compared to the $\ell_{2}$-optimized dither.

Fig.~\ref{fig:lenaqualityentropy} plots the Pareto front for the rate-distortion tradeoff when evaluating SD. The proposed $\ell_{1}$-optimized dither achieves an improved PIQUE at certain low-entropy values for close-range evaluation, while achieving a clearly dominant performance over the $\ell_{2}$-optimized dither for all entropy-PIQUE value pairs at far-range.

\vspace{-2mm}
\Section{Conclusion}
\vspace{-2mm}
We have proposed a theoretical basis for designing distortion-controlled dithered quantization systems and experimentally verified the efficacy of a novel dither distribution in improving compressed image quality at low rates, especially for recompression. These experimental results corroborate provided analytical entropy calculations. Our hypothesis relating quantization error perception to its autocorrelation was validated. Future work includes application of our method to audio or video compression and extension of our framework to dithering non-uniform or vector quantizers.

\Section{References}
\bibliographystyle{IEEEbib}
\bibliography{refs}


\end{document}